\documentclass{sig-alternate-10pt}
\usepackage[margin=1in]{geometry}

\usepackage{url}
\usepackage{graphicx}
\usepackage{booktabs} 
\usepackage{multirow}
\usepackage{todonotes}

\newlength{\figwidths}
\newlength{\expwidths}
\newlength{\expwidthd}
\begin{document}
\pagestyle{empty}

\title{Event Data Quality: A Survey}

\author{Ruihong Huang\\
Tsinghua University \\
hrh16@mails.tsinghua.edu.cn
\and Jianmin Wang\\
Tsinghua University \\
jimwang@tsinghua.edu.cn}

\maketitle

\begin{abstract}

Event data are prevalent in diverse domains such as financial trading, business workflows and industrial IoT nowadays.
An event is often characterized by several attributes denoting the meaning associated with the corresponding occurrence time/duration.
From traditional operational systems in enterprises to online systems for Web services,
event data is generated from physical world uninterruptedly.
However, due to the variety and veracity features of Big data, 
event data generated from heterogeneous and dirty sources could have very different event representations and data quality issues.
In this work, we summarize several typical works on studying data quality issues of event data, including:
(1) event matching, 
(2) event error detection,
(3) event data repairing, and
(4) approximate pattern matching.
\end{abstract}


\section{Introduction}\label{sec:introduction}
Event data generated by large-scale event systems are now prevalent in various domains, 
including financial trading, business process management, network monitoring, supply chain management, social network and health care.
Conventionally, an event is composed of several attributes denoting the meaning and occurrence time/duration of the event, 
such as an (event name, timestamp) pair like (A,1).

\subsubsection*{What Characterizes Event Data}

Event data usually appears with timestamps, in the form of sequences or streams.
Time series can be also seen as a special case of event data.
Implicate complex connections often exist among events, 
and the event data can be represented as graphs or trees to express such relations.
Due to its prevalence, event data is known as typical big data, which constitute a large scale of data assets of the company.

\subsubsection*{Typical Event Data}

Typical event data involved in current researches focus on the following three aspects.
The first category is workflow execution logs.
There are a large amount of event records in OA and ERP systems,
such as the execution logs of a major bus manufacturer, 
recoding the actual productive process in manufacture.
The second category is sensor readings.
Alert events often come from the equipments monitoring specific data,
like water temperature.
If the sensor capture an abnormal reading, it will trigger alert events.
The last category is the event data from big data system logs.
This type of event data often appears in a log-structured, 
such as the task submission event in Cassandra with several information indicating the status of the cluster.


\begin{table*}
\centering
\caption{Summary of the Discussed Methods} 
\label{table-all-methods}
\begin{tabular}{l|l|l}
\hline
\textsf{Research Topic} & \multicolumn{2}{|c}{\textsf{Key Related Work}}  \\ 
\hline 
\multirow{3}{*}{Event Matching} & Linguistic-based & \cite{DBLP:conf/bpm/DijkmanDG09,DBLP:conf/icde/MelnikGR02,DBLP:conf/icse/NejatiSCEZ07,DBLP:conf/caise/WeidlichDM10}\\ 
\cline{2-3}
& Instance-based & \cite{DBLP:journals/tkde/GaoSZWLZ18, DBLP:journals/tkde/SongGWZWY17, DBLP:conf/sigmod/ZhuSL0Z14, DBLP:conf/icde/ZhuSWYS14} \\
\cline{2-3}
& Specification-based & \cite{DBLP:conf/bpm/DijkmanDG09, DBLP:conf/otm/DongWH014, DBLP:conf/apweb/JinWW12}\\
\hline
\multirow{3}{*}{Event Error Detection} & Temporal Constraints & \cite{DBLP:journals/pvldb/SongC016}\\ 
\cline{2-3}
& Petri Net & \cite{DBLP:journals/pvldb/0001SZL13} \\
\cline{2-3}
& Process Models & \cite{DBLP:journals/is/LeoniMA15}\\
\hline
\multirow{2}{*}{Event Data Repairing} & Sequence-structured Events & \cite{DBLP:journals/pvldb/0001SZL13, DBLP:journals/tkde/0001SZLS16, DBLP:conf/sigmod/SongZWY15}\\ 
\cline{2-3}
& Graph-structured Event & \cite{DBLP:journals/pvldb/SongCY014, DBLP:conf/icde/0001SLZP15}\\
\hline
Approximate Pattern Matching & - & \cite{DBLP:conf/icde/LiG15} \\ 
\hline
\end{tabular}
\end{table*}

\subsubsection*{The Data Quality Issues}

Due to the variety and veracity features of Big data, event data generated from heterogeneous and dirty sources could have very various data quality issues.
For example, 
the events, which log the execution of business processes or workflows, often vary in precision, duration and relevance~\cite{langley1999strategies}.
Different event recording conventions or event erroneous executions are likely observed when the execution of a business process is distributed in multiple companies or divisions.
The corresponding event data collected from a heterogeneous environment involve inconsistencies and errors \cite{DBLP:journals/is/RozinatA08}.

Even in the domains of financial trading like Stock data, where people believed data are reliable, a large amount of inconsistent data are surprisingly observed \cite{DBLP:journals/pvldb/LiDLMS12}. 
According to the study, the accuracy of Stock data in \emph{Yahoo! Finance} is 0.93,
and the reasons for imprecise values include ambiguity in information extraction, granularity mismatch or pure mistake.
For instance, the price of SALVEPAR (SY) in the trading event is misused as the price of SYBASE, which is abbreviated to SY as well in some sources.
Such inaccurate values, e.g., taken as the 52-week low price, may seriously mislead business investment.

The data quality issues of event data may have a great impact on the downstream applications,
such as wild data provenance answers \cite{DBLP:conf/sigmod/SunLDC09},
misleading the aggregation profiling in process data warehousing \cite{DBLP:conf/icde/CasatiCSD07},
or obstructing the discovery of interesting process patterns \cite{DBLP:conf/icde/DingCRTHC08}.
Indeed, the event data quality is essential in process mining, 
and known as the first challenge in the Process Mining Manifesto by the IEEE Task Force on Process Mining \cite{DBLP:conf/bpm/AalstAM11}.


\begin{table}[t]
\caption{Notations}
\label{table:notations}
\centering
\smallskip
\begin{tabular}{rp{2.2in}}
\hline\noalign{\smallskip} Symbol & Description \\ \noalign{\smallskip}
\hline\noalign{\smallskip}
$ed(s_1, s_2)$ & string edit distance of $s_1$ and $s_2$ \\ \noalign{\smallskip}
$\mathcal{L}$ & event log \\ \noalign{\smallskip}
$G(V, E, f)$ & event dependency graph \\ \noalign{\smallskip}
$V$ & event vertex set \\ \noalign{\smallskip}
$E$ & edge set \\ \noalign{\smallskip}
$f$ & labeling function  \\ \noalign{\smallskip}
$\bullet v, v\bullet$ & pre-set/post-set of $v$  \\ \noalign{\smallskip}
$\mathcal{N}(\mathcal{P}, \mathcal{T}, \mathcal{F})$ & Petri net  \\ \noalign{\smallskip}
$\mathcal{P}$ & a finite set of places  \\ \noalign{\smallskip}
$\mathcal{T}$ & a finite set of transitions  \\ \noalign{\smallskip}
$\mathcal{F} $ & a set of directed arcs (flow relation)  \\ \noalign{\smallskip}
\hline
\end{tabular}
\end{table}

\section{Event Matching}

Heterogeneous events are often generated by the distinct information system developed separately by different divisions in large-scale corporations.
Duplicate events commonly exist in these heterogeneous processes for the same business activities.
For example, an activity of notifying customers after delivery could be recorded as \textit{Send Notification} in one subsidiary while as \textit{Email Customer} in another.
While rules are often employed to match heterogeneous tuples in databases such as 
matching dependencies \cite{DBLP:conf/cikm/SongC09,DBLP:journals/tkdd/WangSCYC17}
or comparable dependencies \cite{DBLP:journals/vldb/Song0Y13},
they do not consider the temporal information existing in the event data.
The event matching problem is to construct the similarity and matching relationship of events from heterogeneous sources.

\subsection{Linguistic-based}

A straightforward idea of matching events is to compare their names (a.k.a. event labels). 
String edit distance (syntactic similarity) \cite{DBLP:journals/csur/Navarro01} as well as word stemming and the synonym relation \cite{DBLP:conf/aaai/PedersenPM04} or text correlations \cite{DBLP:journals/isci/SongZ014} (semantic similarity) 
are widely used in the label similarity based approaches. 

Dijkman et al. \cite{DBLP:conf/bpm/DijkmanDG09} propose a similarity metric based on graph edit distance for comparing pairs of process models,
to handle the problem of similarity search in process model repositories.
Melnik et al. \cite{DBLP:conf/icde/MelnikGR02} suggest a simple structural algorithm that can be used for matching of diverse data structures.
By converting the models to be matched into directed labeled graphs, 
, an iterative fixpoint computation can be conducted to evaluate what nodes in one graph are similar to nodes in the second graph.
Nejati et al. \cite{DBLP:conf/icse/NejatiSCEZ07} present an approach to matching and merging hierarchical Statecharts models that exploits both
structural and semantic information in the models, 
and ensures that behavioral properties are preserved.
Weidlich et al. \cite{DBLP:conf/caise/WeidlichDM10} propose a re-usable framework ICoP for identifying correspondences between activities in one process and equivalent activities in a similar process,
while taking into account that equivalent activities may be modelled at different levels of granularity, have different labels, and have different control-flow relations to other activities.
In terms of label similarity, the basic idea of the above work is the similarity metric based on string edit distance.


\paragraph*{Notation}
Let $s_1$ and $s_2$ be two strings and $|s|$ be the length of a string $s$. 
The string edit distance $ed(s_1, s_2)$ between $s_1$ and $s_2$ is the minimal number of atomic string operations needed to transform $s_1$ into $s_2$ or vice versa. 
The atomic string operations include inserting a character, deleting a character or substituting a character for another.
The similarity of $s_1$ and $s_2$ is:
\begin{equation}
\nonumber
Sim(s_1, s_2) = 1.0 - \frac{ed(s_1, s_2)}{max(|s_1|, |s_2|)}
\end{equation}

\paragraph*{Example}
For example, the string edit distance between `Verify invoice' and `Verification invoice' is 7.
Specifically, substituting `y' for `i' costs 1, 
and inserting `cation' costs 6. 
Consequently, the string edit similarity is $1.0 - \frac{7}{20}$.

\subsection{Instance-based} 

Usually, there exist relationships among the event instances in an event log.
By constructing a graph to describe the relationships among events, e.g., the frequency of appearing consecutively in an event log \cite{DBLP:conf/bpm/FerreiraG09},
both structural and typographical similarity can be considered to identify the correspondence among events.


\subsubsection{Matching Similarity}

Simrank \cite{DBLP:conf/kdd/JehW02} like behavioral similarity (BHV) \cite{DBLP:conf/icse/NejatiSCEZ07} considers a global evaluation via propagating similarities in the entire graph in multiple iterations.
Following the same line,
Zhu \cite{DBLP:conf/sigmod/ZhuSL0Z14} and Gao \cite{DBLP:journals/tkde/GaoSZWLZ18} propose a similarity function by iteratively computing neighbor similarities.

\paragraph*{Notation}
An event dependency graph $G$ is a labeled directed graph $\left(V, E, f\right)$, where each vertex in $V$ corresponds to an event, $E$ is an edge set, and $f$ is a labeling function of normalized frequencies that

\noindent
(1) for each $v \in V$, $f(v, v)$ is the normalized frequency of event $v$, i.e., the fraction of traces in the event log $\mathcal{L}$ that contain $v$, and

\noindent
(2) for each edge $(v_1, v_2) \in E$, $f(v_1, v_2)$ is the normalized frequency of two consecutive events $v_1v_2$, i.e., the fraction of traces in which $v_1v_2$ occur consecutively at least once.

\noindent
For any $v \in V$, the pre-set of $v$ is defined as $\bullet v = \{v' | (v', v) \in E\}$ and the post-set of $v$ is defined as $v \bullet = \{v' | (v, v') \in E\}$.

Considering the presence of dislocated matching, 
any event in an event log can be a starting/ending event.
An artificial event $v^X$ is added into $V$, which denotes the virtual beginning/end of all traces in an event log.
For each event $v \in V$ except $v^X$, 
two artificial edges $(v, v^X)$ and $(v^X, v)$ are added, i.e., each event can be a virtual starting event (edge $(v^X, v)$) and a virtual ending event (edge $(v, v^X)$). 
Moreover, the label of the virtual edge is defined as $f(v^X, v)$ = $f(v, v^X)$ = $f(v)$ based on the intuition that a trace can start/end with event $v$ at all the locations where $v$ occurs.

The forward similarity of two events is
\begin{equation}
\nonumber 
\mathcal{S}(v_1, v_2) = \alpha (s(v_1, v_2) + s(v_2, v_1)) / 2 + (1 - \alpha)\mathcal{S}^L(v_1, v_2)
\end{equation}
where $\mathcal{S}^L(v_1, v_2)$ is the label similarity of $v_1$ and $v_2$, $\alpha \in [0, 1]$ is a weight, 
$s(v_1, v_2)$ and $s(v_2, v_1)$ are one-side similarities 
\begin{equation}
\nonumber
s(v_1, v_2) = \frac{\sum_{v_1' \in \bullet v_1} max_{v_2' \in \bullet v_2} C(v_1, v_1', v_2, v_2') \mathcal{S}(v_1', v_2')}{|\bullet v_1|} 
\end{equation}
given that $C(v_1, v_1', v_2, v_2') = c * (1 - \frac{|f(v_1, v_1') - f(v_2, v_2')|}{f(v_1, v_1') + f(v_2, v_2')})$, where $c$ is a constant that has $0 < c < 1$.

\paragraph*{Example}
\begin{figure}[t]
\centering
\includegraphics[width=\figwidths]{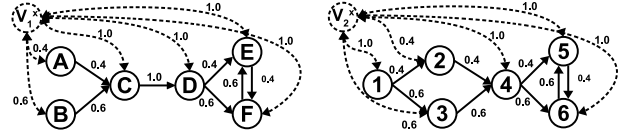}
\caption{Example dependency graph with artificial events}
\label{fig-simrank-exp}
\end{figure}

Consider the two dependency graphs in Figure \ref{fig-simrank-exp}.
At the beginning of the iteration, $\mathcal{S}_0(v_1^X, \\v_2^X)$ is assigned with $1.0$, 
and $\mathcal{S}_0(v_1, v_2)$ is assigned with $0$ for any other event pairs where $v_1 \neq v_1^X$ and $v_2 \neq v_2^X$.
Consider the event pair $(A, 1)$.
Let $\alpha = 1$, on the first iteration, 
we have $s^1(A, 1) = \frac{1}{|\bullet A|}C(v_1^X, A, v_2^X, 1) \mathcal{S}^0(v_1^X, v_2^X) = 0.457$
and $s^1(1, A) \\= \frac{1}{|\bullet 1|}C(v_1^X, 1, v_2^X, A) \mathcal{S}^0(v_1^X, v_2^X) = 0.457$,
so that $\mathcal{S}^1(A, 1) = 0.5 * (s^1(A, 1) + s^1(1, A)) = 0.457$.
For the event pair $(A, 2)$, we have $s^1(A, 2) = \frac{1}{|\bullet A|} max(\\C(v_1^X, A, v_2^X, 2) \mathcal{S}^0(v_1^X, v_2^X), C(v_1^X, A, 1, 2) \mathcal{S}^0(v_1^X, 1)) \\= 0.8$
and $s^1(2, A) = \frac{1}{|\bullet 2|} max(C(v_2^X, 2, v_1^X, A) \mathcal{S}^0(v_2^X, \\v_1^X), C(1, 2, v_1^X, A) \mathcal{S}^0(1, v_1^X)) = 0.4$,
so that $\mathcal{S}^1(A, 2) \\= 0.5 * (0.8 + 0.4) = 0.6$.
An average similarity of all the event pairs can be computed likewise.

\subsubsection{Mapping Correspondence}

However, sometimes event names could be opaque (e.g.,merely with obscure IDs), and the structure based matching techniques may also fail to perform owing to the poor discriminative power of dependency relationships between events.
Owing to the absence of typographic or linguistic similarity,
normal distance for matching with opaque names (OPQ) \cite{DBLP:conf/sigmod/KangN03} concerns a local evaluation of similar neighbors. 
Considering more complex event patterns often exist in event logs and may serve as more discriminative features,
Zhu \cite{DBLP:conf/icde/ZhuSWYS14} and
Song \cite{DBLP:journals/tkde/SongGWZWY17} propose a generic pattern based matching framework, which is compatible with the existing structure based techniques.

\paragraph*{Notation}
Let $M$ be any mapping of vertices (events) over the two dependency graphs $G_1(V_1, E_1, f_1)$ and $G_2(V_2, E_2, f_2)$. 
A score function is employed w.r.t. the mapping $M$, 
namely \emph{normal distance}, 
to evaluate the similarity of two event logs.
The normal distance $D^N(M)$ of $M$ is defined as
\begin{equation}
\nonumber
\sum_{v_1, v_2 \in V_1}\left(1 - \frac{|f_1(v_1, v_2) - f_2(M(v_1), M(v_2))|}{f_1(v_1, v_2) + f_2(M(v_1), M(v_2))}\right)
\end{equation}
Two forms of normal distances are studied. 
If $v_1 = v_2$ is required in the formula, 
the normal distance considers only the frequencies of individual events, 
i.e., \emph{vertex form}. 
Otherwise, the normal distance is in \emph{vertex+edge} form which considers both vertex frequencies and edge frequencies.

Normal distance is the summation of frequency similarities (or differences) of corresponding vertices or edges w.r.t. mapping $M$. 
The higher the normal distance is, the more similar the vertices and edges captured by $M$ are.
Consequently, the matching problem is to find a mapping $M$ that has the highest normal distance.

The complex patterns with SEQ and AND operators can be discriminative features in event matching.
An event pattern specifies particular orders of event occurrence, which are defined recursively:
\begin{itemize}
\item A single event e is an event pattern;
\item  SEQ($p_1, p_2, \dots, p_k$) is an event pattern in which the patterns $p_i, i \in 1, \dots, k$, occur sequentially;
\item AND($p_1, p_2, \dots, p_k$) is an event pattern that requires the concurrent occurrence of the patterns $p_i, i \in 1, \dots, k$, i.e., the order of $p_i$ does not matter.
\end{itemize}
An event pattern can naturally be represented as a directed graph, where each vertex corresponds to an event \cite{DBLP:conf/sigmod/WuDR06}. 
Intuitively, SEQ operator specifies edges between consecutive $p_i$ and $p_{i+1}, i \in 1, \dots, k - 1$, 
while AND operator indicates edges between any two $p_i$ and $p_j, i \neq j, i, j \in 1, \dots, k$.
It is worth noting that for all the events $e_1, e_2, \dots, e_k$ included in a pattern and $i \neq j, i, j \in 1, \dots, k$,
it is assumed that there should be $e_i \neq e_j$, 
since some translated graphs of distinct patterns may be the same if the duplication of events are permitted
(e.g., SEQ$(A, B, A, B)$ and AND$(A, B)$).

Let $M$ be a mapping of events over dependency graphs $G_1(V_1, E_1, f_1)$ and
$G_2(V_2, E_2, f_2)$. 
For a set of patterns $\mathcal{P}$, the pattern normal distance of $M$ is defined as
\begin{equation}
\nonumber
D^N(M)=\sum_{p \in \mathcal{P}}\left(1 - \frac{|f_1(p) - f_2(M(p))|}{f_1(p) + f_2(M(p))}\right)
\end{equation}
, where $M(p)$ is the pattern in $G_2$ corresponding to $p$ in $G_1$ via the mapping $M$ such that each event $v$ in $p$ maps to an event $M(v)$ in $M(p)$.

\paragraph*{Example}
\begin{figure}[t]
\centering
\includegraphics[width=\figwidths]{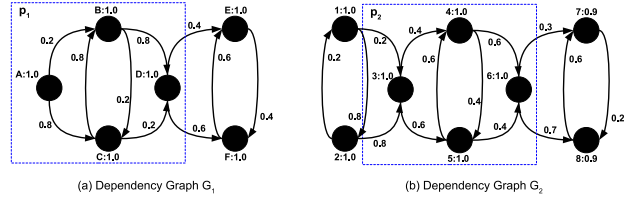}
\caption{Example dependency graph with patterns}
\label{fig-pattern-exp}
\end{figure}

Consider a pattern $p_1$= SEQ($A$, AND($B$, $C$), $D$) in Figure \ref{fig-pattern-exp}(a) from \cite{DBLP:journals/tkde/SongGWZWY17}. 
The pattern $p_1$ is illustrated as a graph.
The vertices of events are {$A, B, C, D$}. 
Two edges $BC, CB$ are added due to pattern AND($B$, $C$). 
According to SEQ($A$, AND($B$, $C$), $D$), both $B$ and $C$ can be performed after $A$ 
and should be done before $D$. 
Thus, another 4 edges $AB, AC, BD$ and CD are added. 
The graph translated from $p_1$ is a subgraph of $G_1$ surrounded by blue dashed line in Figure \ref{fig-pattern-exp}(a).

For the true mapping $M = \{A \rightarrow 3, B \rightarrow  4, C \rightarrow  5, D \rightarrow 6, E \rightarrow 7, F \rightarrow 8\}$, 
pattern $p_1$ (in $G_1$) corresponds to a subgraph $p_2$ in $G_2$. 
Since all traces in $\mathcal{L}_1$ and $\mathcal{L}_2$ match with $p_1$ and $p_2$, respectively, 
we have $f_1(p_1) = f_2(p_2) = 1.0$. 
By considering all vertices and edges as patterns in the above Formula, 
the pattern normal distance of M is $D^N(M) = 14.91$.

However, the pattern $p_1$ has no mapped pattern w.r.t. $M' = \{A \rightarrow 6, B \rightarrow 2, C \rightarrow 1, D \rightarrow 3, E \rightarrow 4, F \rightarrow 5\}$. 
The pattern normal distance of $M'$ is 14. 
While without $p_1$, $M'$ has a higher score with $D_v^N(M') > D_v^N(M) $ and  $D_{v + e}^N(M') > D_{v + e}^N(M)$. 
By introducing $p_1$, the true mapping M with the highest pattern normal distance beats $M'$.

\subsection{Specification-based} 
There are many other specification-based event matching researches.
Dijkman et al. \cite{DBLP:conf/bpm/DijkmanDG09} conduct graph matching algorithms for business process model similarity search.
Dong et al. \cite{DBLP:conf/otm/DongWH014} focus on a behavioral process similarity algorithm named CFS based on complete firing sequences which are used to express model behavior.
Since dynamic behavior is the essential characteristic of workflow models, 
Jin et al. \cite{DBLP:conf/apweb/JinWW12} measure the similarity between models based on their behavior.

\section{Event Error Detection} 
Due to the complex circumstances in the real word,
the event recorded in the log could exist various types of errors.
While integrity constraints such as 
differential dependencies \cite{DBLP:journals/tods/Song011,DBLP:journals/tkde/Song0C14}, 
are often employed to detect the errors in databases, 
they are not directly applicable to the event data with complicated temporal relationships.
Fortunately, prior knowledge could be used in figuring out such imprecise data.
The event error detection problem is to find out the erroneous instances that contradict the predefined constraints.

\subsection{Temporal Constraints}
\label{sect-Temporal-Constraint}

Dechter et al. \cite{DBLP:journals/ai/DechterMP91} introduce \emph{temporal constraint satisfaction problem} (TCSP),
where variables represent time points and temporal information is represented by a set of unary or binary constraints, each specifying a set of permitting intervals.

\paragraph*{Notation}
A temporal constraint can be presented by a set of intervals:
$$ \{[\mathit{a}_1, \mathit{b}_1], \dots, [\mathit{a}_n, \mathit{b}_n]\}. $$
According to the different types of the constraint, it represents different meanings:
a unary constraint $\mathit{T}_i$ restricting the domain of variable $\mathit{X}_i$ represents the disjunction
$$(\mathit{a}_1 \leq \mathit{X}_i \leq \mathit{b}_1) \vee \dots \vee (\mathit{a}_n \leq \mathit{X}_i \leq \mathit{b}_n),$$
while a binary constraint $\mathit{T}_{ij}$ constraints the value for distance $\mathit{X}_i - \mathit{X}_j$, which represents the disjunction
$$(\mathit{a}_1 \leq \mathit{X}_i - \mathit{X}_j \leq \mathit{b}_1) \vee \dots \vee (\mathit{a}_n \leq \mathit{X}_i - \mathit{X}_j \leq \mathit{b}_n),$$

A  temporal constraint satisfaction problem involves a set of variables, $\mathit{X}_1, \dots, \mathit{X}_n$,
having continuous domains, and each variable represents a time point.
A special time point $\mathit{X}_0$ represents the beginning of the world, and all times are relative to $\mathit{X}_0$.
Therefore, a unary constraint $\mathit{X}_i$ could be treated as a binary constraint $\mathit{X}_{0i}$.
The solution of the problem is a tuple $\mathit{X} = (\mathit{x}_1, \dots, \mathit{x}_n)$ satisfying all the constraints.

\paragraph*{Example}

\begin{figure}[t]
\centering
\includegraphics[width=0.8\figwidths]{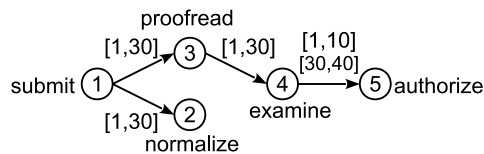}
\caption{An example temporal constraint network}
\label{fig-temporal-event}
\end{figure}

Figure \ref{fig-temporal-event} presents an example temporal constraint network from  \cite{DBLP:journals/pvldb/SongC016}. 
It denotes the steps (a.k.a. events, denoted by nodes) that are required in every part design process of a train manufacturer. 
The temporal network (abstracted from workflow specifications) specifies the constraints on occurring timestamps of events. 
For instance, the temporal constraint $[1,30]$ from event 1 (\textsf{submit}) to event 3 (\textsf{proofread}) indicates the minimum and maximum restrictions on the distance (delay) of these two events' timestamps. 
That is, event 3 (\textsf{proofread}) should be processed within 30 minutes after event 1 (\textsf{submit}).
Multiple intervals may also be declared between two events. 
For instance, $[1,10],[30,40]$ on edge $4\rightarrow5$ denote that event 5 (\textsf{authorize}) can be processed after event 4 (\textsf{examine}) either by the department head in 1-10 minutes or by the division head in 30-40 minutes. 

Consider a corresponding example instance of event trace in Table \ref{table-temporal-event}.
It records one instance of five steps (events) for processing a part design work, including \textsf{submit, normalize, proofread}, etc.
Each event is associated with a timestamp on when this event occurred. 
Events 3 and 1 in $t_3$ and $t_1$ satisfy the temporal constraint, since their timestamp distance $09{:}25-09{:}05=20$ is in the range of $[1,30]$.
Note that the events are collected from various external sources. 
Imprecise timestamps are prevalent, e.g., {23:53} of event 2 (\textsf{normalize}) in $t_5$, which is delayed until just before midnight owing to latency.
The imprecise timestamps are identified as violations of the temporal constraints, such as events 2 and 1 with timestamp distance $23{:}53-09{:}05=888>30$.

 \begin{table}
 \centering\small
 \caption{An example relation instance of events} 
 \label{table-temporal-event}
 \begin{tabular}{lll}
 \toprule
 & \textsf{event} & \textsf{timestamp}  \\ 
 \midrule 
 $t_1$ & 1 (\textsf{submit})    & 09:05 \\ 
 $t_2$ & 5 (\textsf{authorize}) & 09:54 \\ 
 $t_3$ & 3 (\textsf{proofread}) & 09:25 \\ 
 $t_4$ & 4 (\textsf{examine})   & 09:48 \\ 
 $t_5$ & 2 (\textsf{normalize}) & \textbf{23:53} \\ 
 \bottomrule
 \end{tabular}
 \end{table}



%

%

\paragraph*{Application}
Based on temporal constraints, Dechter et al. \cite{DBLP:journals/ai/DechterMP91} present algorithms for performing the following reasoning tasks:
(1) finding all feasible times that a given event can occur;
(2) finding all possible relationships between two given events;
(3) generating one or more scenarios consistent with the information provided.
Song et al. \cite{DBLP:journals/pvldb/SongC016}
utilize the temporal constraints to repair the erroneous timestamps.

\subsection{Petri Net}
\label{sect-Petri-Net}

\begin{figure}[t]
\centering
\includegraphics[width=\figwidths]{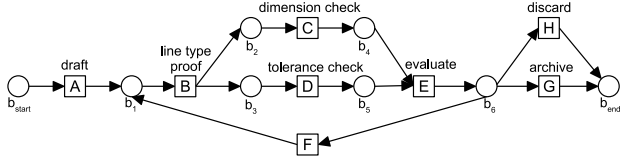}
\caption{Example Petri net}
\label{fig:petri_net}
\end{figure}

Petri net \cite{24143} is a graphical and mathematical modeling tool.
It is a promising tool for describing and studying information processing systems.
As a graphical tool, Petri net can be used as a visual-communication aid
similar to flow charts and networks, for presenting a specific process.
Moreover, Petri net can be also considered as constraints for data repairing task.

\paragraph*{Notation}
A Petri net is a triplet $\mathcal{N}(\mathcal{P}, \mathcal{T}, \mathcal{F})$,
where $\mathcal{P}$ is a finite set of places, $\mathcal{T}$ is a finite set of transitions,
and $\mathcal{F} \subseteq (\mathcal{P} \times \mathcal{T}) \cup (\mathcal{T} \times \mathcal{P})$
is a set of directed arcs (flow relation).
A process specification is a Petri net $\mathcal{N}_s(\mathcal{P}_s, \mathcal{T}_s, \mathcal{F}_s)$
which has a unique source place $\mathsf{b}_{\mathsf{start}} \in \mathcal{P}_s$, whose pre-set is empty,
and a unique sink place $\mathsf{b}_{\mathsf{end}} \in \mathsf{P}_s$, whose post-set is empty.
Each node $\mathit{x} \in \mathcal{P}_s \cup \mathcal{T}_s$ is on a path
from $\mathsf{b}_{\mathsf{start}}$ to $\mathsf{b}_{\mathsf{end}}$.

\paragraph*{Example}
Figure \ref{fig:petri_net} presents a process specification for producing an engineering drawing in a train manufacturer.
Each square denotes a transition and each circle denotes a place.
All the arrows attached to a transition denote the corresponding flows should be executed in parallel.
For instance, both task $\mathsf{C}$ and task $\mathsf{D}$ should be conducted after task $\mathsf{B}$.
While only one of the flows going out a place can be executed.
For instance, $\mathsf{b}_6$ leads to either task $\mathsf{F}$, task $\mathsf{G}$, or task $\mathsf{H}$ after task $\mathsf{E}$.

\paragraph*{Application}
Petri nets are directly employed in a number of real applications,
for modeling personnel management processes, biological information, and workflows.
Wang et al. \cite{DBLP:journals/pvldb/0001SZL13} study the missing event data recovering problem
which use petri nets as constraints.
They propose efficient techniques to find the minimum recovery of missing data that meets the contraints.


\subsection{Process Models}

In the context of process mining,
conformance checking techniques verify whether the observed behavior recorded in an event log matches a modeled behavior.
This type of analysis is crucial in many domains such as process auditing and risk analysis, 
since the actual process executions often deviate from the predefined theoretical models.
Leoni et al. \cite{DBLP:journals/is/LeoniMA15} propose  an alignment-based framework to check the conformance of declarative process models and to preprocess event-log data.
The event logs are aligned declarative models, i.e., events in the log are related to activities in the model if possible. 
The alignment also provides sophisticated diagnostics that pinpoint where deviations occur and how severe they are.

\section{Event Data Repairing}

\subsection{Sequence-structured Events} 

\subsubsection{Under Process Specification Constraints}

Existing approaches \cite{DBLP:conf/bpm/LeoniMA12} on cleaning event data  treat event logs as unstructured sequences.
The minimal recovery of missing events is also studied as optimal sequence alignment \cite{DBLP:conf/ijcai/KobayashiKW11}.
Leoni et al. \cite{DBLP:conf/bpm/LeoniMA12} and Adriansyah et al. \cite{DBLP:conf/socialcom/AdriansyahDZ13} developed alignment-based approaches to repair event logs w.r.t. declarative models as specifications.
Beside inserting missing events, they also consider removing the existing events \cite{DBLP:conf/bpm/LeoniMA12} and swapping two events \cite{DBLP:conf/socialcom/AdriansyahDZ13} to repair the event logs.
$A^*$ algorithms are employed in these works.
The basic idea is to enumerate all the valid combinations of events as possible sequences, and apply the $A^*$ algorithm to search the one with the minimum cost.
The alignment approach considers a search space involving redundant sequences with respect to parallel events.
Wang et al. propose to recover missing events with process model constraints \cite{DBLP:journals/pvldb/0001SZL13, DBLP:journals/tkde/0001SZLS16}.
The proposed approach can successfully avoid such inefficient scenario and show significantly lower time cost in the experiments.

\paragraph{Notation}
Based on the concept of Petri Net, a firing sequence of a process specification $\mathcal{N}(\mathcal{P}, \mathcal{T}, \mathcal{F})$, 
and its post-set, are defined recursively as follows: 
\begin{enumerate}
\item The empty sequence $\epsilon$ is a firing sequence, and $\epsilon \bullet = {b_{start}}$;
\item If $\sigma$ is a firing sequence, $e \in \mathcal{T}_s$ is a transition (event), 
and $\bullet e \subseteq \sigma \bullet$, 
then $\sigma e$ is also a firing sequence, 
and $(\sigma e)\bullet = (\sigma \bullet) - (\bullet e) + (e \bullet)$.
\end{enumerate}
A sequence $\sigma$ is said conforming to a process specification,
denoted by $\sigma \models \mathcal{N}_s$,
if $\sigma$ is a firing sequence w.r.t. $\mathcal{N}_s$ and $\sigma \bullet = {b_{end}}$.

Let $\sigma$ be a firing sequence.
For the next event (transition) $e$, 
if $\bullet e \not \subseteq \sigma_k \bullet$,
we call $(\sigma_k \bullet, \bullet e)$ a gap with at least one missing event between $\sigma_k$ and $e$.

A gap indicates that the previous firing sequence $\sigma$ is successfully executed so far
and it is impossible to execute the next event $e$ further.
In other words, $\sigma e$ is not a firing sequence.
There are some events missing between $\sigma$ and $e$.

For a gap $(\sigma \bullet, \bullet e)$, 
we call a transition (event) sequence $\tau \in \mathcal{T}^*$ a fill of the gap, if it ensures
\begin{enumerate}
	\item $\sigma \tau$ is a firing sequence,
	\item $\bullet e \subseteq (\sigma \tau)\bullet$.
\end{enumerate}

\paragraph{Example}
\begin{figure}[t]
\centering
\includegraphics[width=\figwidths]{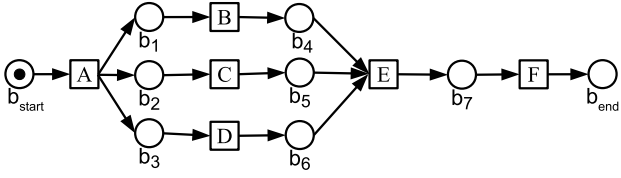}
\caption{Example process specification}
\label{fig-recover-exp}
\end{figure}
To recover a sequence $\langle AEF\rangle$ over the process specification in Figure \ref{fig-recover-exp}, 
the existing aligning approach enumerates all possible combinations of events in parallel, i.e., $\langle ABCDEF\rangle$, $\langle ABDCEF\rangle$, $\langle ACBD\\EF\rangle$, $\langle ACDBEF\rangle$,
$\langle ADBCEF\rangle$ and $\langle ADCBEF\rangle$. 
However, any topological sort should always be a minimum recovery, 
e.g., $\langle ABCDEF\rangle $ with the minimum distance $3$ to the input sequence $\langle AEF\rangle $. 
There is no need to enumerate other redundant recoveries.

To recover the gap between a firing sequence $\langle A\rangle $ and event $E$, 
the program generates a set of places $X = \{b_4, b_5, b_6\}$ in $\bullet E$,
but not in $\langle A\rangle \bullet = \{b_1, b_2, b_3\}$.
For each event in $\bullet X = \{B,C,D\}$, e.g., $B$, we fill the gap between the current firing sequence $\langle A\rangle \bullet$ and $\bullet B$. 
It outputs a firing sequence $\langle AB\rangle $ with post-set $\langle AB\rangle \bullet$ = $\{b_4, b_2, b_3\}$. 
Next, by inserting $C \in \bullet X$, we have $\langle ABC\rangle \bullet = \{b_4, b_5, b_3\}$. 
It follows $\langle ABCD\rangle \bullet = \{b_4, b_5, b_6\}$. 
Finally, the gap between $\langle A\rangle $ and $E$ is filled by $\langle BCD\rangle $ such that $\bullet E \subseteq \langle ABCD\rangle \bullet$.


\subsubsection{Under Speed Constraints}
When the contents of events appear in an numerical form,
there are more abundant relationships among the event points.
Song et al. \cite{DBLP:conf/sigmod/SongZWY15} propose to clean stream data under speed constraints.

\paragraph{Notation}
Consider a sequence $\mathit{x}=\mathit{x}[1], \mathit{x}[2], \dots$, where each $\mathit{x}[i]$ is the value of the $i$-th data point. 
Each $\mathit{x}[i]$ has a timestamp $\mathit{t}[i]$.
For brevity, we write $\mathit{x}[i]$ as $\mathit{x}_i$, and $\mathit{t}[i]$ as~$\mathit{t}_i$.

A \emph{speed constraint} $\mathit{s}=(\mathit{s}_{\min},\mathit{s}_{\max})$ with window size $\mathit{w}$ is a pair of minimum speed $\mathit{s}_{\min}$ and maximum speed $\mathit{s}_{\max}$ over the sequence $\mathit{x}$.
We say that a sequence $\mathit{x}$ \emph{satisfies} the speed constraint $\mathit{s}$, denoted by $\mathit{x}\vDash\mathit{s}$, if for any $\mathit{x}_i, \mathit{x}_j$ in a window, i.e.,  
$0<\mathit{t}_j-\mathit{t}_i\leq\mathit{w}$, it has 
$\mathit{s}_{\min}\leq\frac{\mathit{x}_j-\mathit{x}_i}{\mathit{t}_j-\mathit{t}_i}\leq\mathit{s}_{\max}.$

The window $w$ denotes a period of time. 
In real settings, speed constraints are often meaningful  within a certain period. 
For example, it is reasonable to consider the maximum walking speed in hours (rather than the speed between two arbitrary observations in different years), since a person usually cannot keep on walking in his/her maximum speed for several years without a break. 
In other words, it is sufficient to validate the speed w.r.t.\ two points $\mathit{x}_i,\mathit{x}_j$ in a window $w=24$ hours, i.e., whether $\mathit{s}_{\min}\leq\frac{\mathit{x}_j-\mathit{x}_i}{\mathit{t}_j-\mathit{t}_i}\leq\mathit{s}_{\max}, 0<\mathit{t}_j-\mathit{t}_i\leq\mathit{w}$. 
In contrast, considering the speed w.r.t.\ two points in an extremely large period (e.g., two observation points in different years) is meaningless and unnecessary. 
Similar examples include the speed constraints on stock price whose daily limit is directly determined by the price of the \emph{last} trading day, i.e., with window size 1. 


The speed constraint $\mathit{s}$ can be either positive (restricting value increase) or negative (on decrease).
In most scenarios, the speed constraint is natural, e.g., the fuel consumption of a crane should not be negative and not exceed 40 liters per hour, while some others could be derived. 

A \emph{repair} $\mathit{x}'$ of $\mathit{x}$ is a modification of the values $\mathit{x}_i$ to $\mathit{x}'_i$ where $\mathit{t}'_i=\mathit{t}_i$.
Referring to the minimum change principle in data repairing \cite{DBLP:conf/sigmod/BohannonFFR05},
the repair distance is evaluated by the difference between the original $\mathit{x}$ and the repaired $\mathit{x}'$,
\begin{align*}
\label{equation:repair-cost}
\Delta(\mathit{x},\mathit{x}')=\sum_{\mathit{x}_i\in\mathit{x}}|\mathit{x}_i-\mathit{x}'_i|.
\end{align*}

\paragraph{Example}
\begin{figure}[t]
\centering
\includegraphics[width=\figwidths]{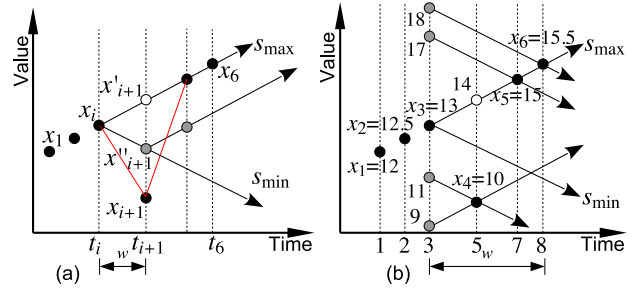}
\caption{Possible repairs under speed constraints}
\label{fig-speed-exp}
\end{figure}

Consider a sequence $\mathit{x} = \{12, 12.5, 13, 10, 15, 15.5\}$ of six data points, with timestamps $\mathit{t} = \{1,2,3,5,7,8\}$. 
Figure \ref{fig-speed-exp}(a) from \cite{DBLP:conf/sigmod/SongZWY15} illustrates the data points (in black). 
Suppose that the speed constraints are $\mathit{s}_{\max} = 0.5$ and $\mathit{s}_{\min} = -0.5$.

For a window size $\mathit{w} = 2$ in the speed constraints,  data points $\mathit{x}_3$ and $\mathit{x}_4$, with timestamp distance $5-3\leq 2$ in a window, are identified as violations to $\mathit{s}_{\min} = -0.5$, since the speed is
$\frac{10-13}{5-3}=-1.5<-0.5$.
Similarly, $\mathit{x}_4$ and $\mathit{x}_5$ with speed 
$\frac{15-10}{7-5}=2.5>0.5$ are violations to $\mathit{s}_{\max} = 0.5$.

To remedy the violations (denoted by red lines), a repair on $\mathit{x}_4$ can be performed, i.e., $\mathit{x}'_4=14$ (the white data point). 
As illustrated in Figure \ref{fig-speed-exp}(a), the repaired sequence satisfies both the maximum and minimum speed constraints.
The repair distance is $\Delta(\mathit{x},\mathit{x}')=|10-14|=4$.

Note that if the window size is too small such as $\mathit{w} = 1$, the violations between  $\mathit{x}_3$ and $\mathit{x}_4$ (as well as $\mathit{x}_4$ and $\mathit{x}_5$) could not be detected, since their timestamp distance is $2>1$.
On the other hand, if the window size is too large, say $\mathit{w} = 10$, then all the pairs of data points in $\mathit{x}$ have to be compared.
Although the same repair $\mathit{x}'$ is obtained, the computation overhead is obviously higher (and unnecessary). 
Song et al. \cite{DBLP:conf/sigmod/SongZWY15} also propose to determine an adaptive window size for balancing accuracy and efficiency.

\subsection{Graph-structured Events} 

It is worth noting that structural information do exist among events.
A very common example is the task passing relationships, e.g., the manager assigns the work to another staff for succeeding operations.
Such structural information are not only essential to obtaining more precise event repairs but also useful in improving the computation efficiency.

\subsubsection{Under Neighborhood Constraint}
Neighborhood constraints, specifying label pairs that are allowed to appear on adjacent vertexes in the graph, 
are employed to detect and repair erroneous vertex labels.
Song et al. \cite{DBLP:journals/pvldb/SongCY014} propose to repair the event data with neighborhood constraints specified by simple graphs.

\paragraph*{Notation}
\begin{figure}[t]
\centering
\includegraphics[width=\figwidths]{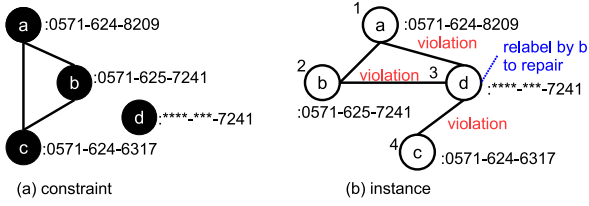}
\caption{Example of neighborhood constraint}
\label{fig-neighbor-not}
\end{figure}

Two labels $l_1, l_2$ match a constraint graph $\mathcal{S}(L, N)$, 
denoted by $(l_1, l_2) \asymp \mathcal{S}$, 
if either $l_1 = l_2$ denotes the same label or $(l_1, l_2) \in N$ is an edge in $\mathcal{S}$. 
For example, in Figure \ref{fig-neighbor-not}(a), 
we have $(a, a) \asymp \mathcal{S}, (a, b) \asymp \mathcal{S}$, but $(a, d) \not\asymp \mathcal{S}$. 
It is worth noting that $(a, a) \asymp \mathcal{S}$ implies the self-loop relationship of labels in S.

An instance graph $G(V, E)$ satisfies a constraint graph $\mathcal{S}(L, N)$, 
denoted by $G \models S$, if $(v, u) \in E$ implies $(\lambda(v), \lambda(u)) \asymp \mathcal{S}, \forall(v, u) \in E$. 
That is, for any edge $(v, u) \in E$, 
their labels $\lambda(v), \lambda(u)$ must match the constraint graph $\mathcal{S}$ with either $\lambda(v) = \lambda(u)$ or $(\lambda(v), \lambda(u)) \in N$.

We call $(v, u)$ a violation to the constraint graph $\mathcal{S}$, 
if $(v, u) \in E$ and $(\lambda(v), \lambda(u)) \not\asymp S$. 
For example, in Figure \ref{fig-neighbor-not}(b), 
the edge $(1, 3)$ indicates a violation to $\mathcal{S}$, 
as their labels $(a, d) \not\asymp S$ are neither the same nor adjacent in Figure \ref{fig-neighbor-not}(a) of constraints.

\paragraph*{Example}
\begin{figure}[t]
\centering
\includegraphics[width=\figwidths]{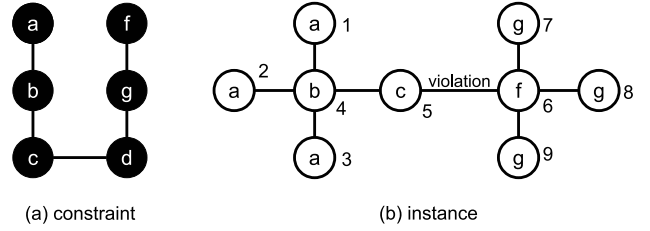}
\caption{Example of repairing under neighborhood constraint}
\label{fig-neighbor-exp}
\end{figure}

Consider the example in Figure \ref{fig-neighbor-exp} with relabeling cost $\delta(c, g) = \delta(f, d) = 1$. 
When conducting a greedy method, the best choice may be repairing vertex $6$ with label $d$. 
The violation between vertexes $5$ and $6$ is eliminated, 
and there is no new violation among vertexes introduced after relabeling vertex $6$ to $d$. 
The relabeling steps terminate with all the neighborhood constraints satisfied.

\subsubsection{Under Process Model Constraints}
The structural information existing among events are often expressed in the form of process model as well,
like the designing process model in the bus manufacture company.
Wang et al. \cite{DBLP:conf/icde/0001SLZP15} propose the event data repairing method with process model constraints. 

\paragraph*{Notation}
A causal net is a Petri net $N = (P,T, F)$ such that for every $p \in P$, $|pre_F(p)| \leq 1$ and $|post_F(p)|\\ \leq 1$.

It is easy to see that there will be no XOR-split or XOR-join in a causal net 
(according to the maximum in/out degree 1 of places), 
while AND-split and AND-join are allowed. 
If we interpret places as edges connecting two transitions, the net is indeed a directed acyclic graph of transitions.

An execution of a process specification $N_s(P_s, T_s, F_s)$ is denoted by $(N_\sigma, \pi)$, 
where $N_\sigma(P_\sigma,T_\sigma, F_\sigma)$ is a causal net and 
$\pi$ is a labeling $\pi: P_\sigma \cup T_\sigma \rightarrow P_s \cup T_s$ such that $\pi(P_\sigma) \subseteq P_s$, and $\pi(T_\sigma) \subseteq T_s$.

We say an execution $(N_\sigma, \pi)$ conforms to a process specification $N_s$, 
denoted by $(N_\sigma, \pi) \models N_s$, 
if and only if 
(i) $\pi(P_\sigma) \subseteq P_s$ and $\pi(T_\sigma) \subseteq T_s$; 
(ii) for any $t \in T_\sigma$, $\pi(pre_{F_\sigma} (t)) = pre_{F_s} (\pi(t))$ and $\pi(post_{F_\sigma} (t)) = post_{F_s} (\pi(t))$; 
(iii) for any $p \subseteq P_\sigma$, $pre_{F_\sigma} (p) = \emptyset$ implies $pi(p)$ = start and $post_{F_\sigma} (p) = \emptyset$ indicates $\pi(p)$ = end.

\paragraph*{Example}
\begin{figure}[t]
\centering
\includegraphics[width=\figwidths]{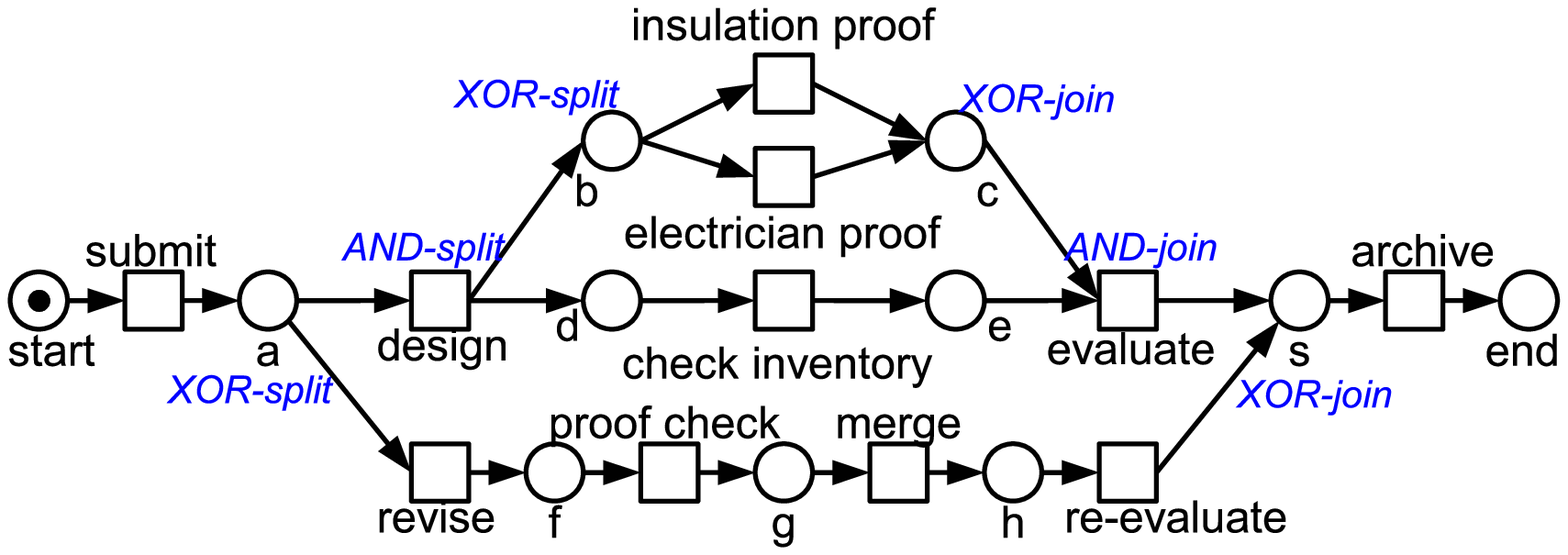}\\
(a) Specification for \textsf{part design} process\\
\begin{tabular}{llll}
\hline\noalign{\smallskip} \textbf{Event} & \textbf{Name} & \textbf{Operator} & \textbf{Successor}\\
\hline \noalign{\smallskip}
$t_1$ & \textsf{submit}  &  A &  B \\ \noalign{\smallskip}
$t_2$ & \textsf{design}  &  B & C \& D  \\ \noalign{\smallskip}
$t_3$ & \textsf{insulation proof}  &  C & E    \\ \noalign{\smallskip}
$t_4$ & \textsf{check inventory}  &  D & E  \\ \noalign{\smallskip}
$t_5$ & \textsf{evaluate}  &  E & F    \\ \noalign{\smallskip}
$t_6$ & \textsf{archive}  &  F & ---    \\ \noalign{\smallskip}
\hline
\end{tabular}\\
(b) Example of an execution trace\\
\includegraphics[width=\figwidths]{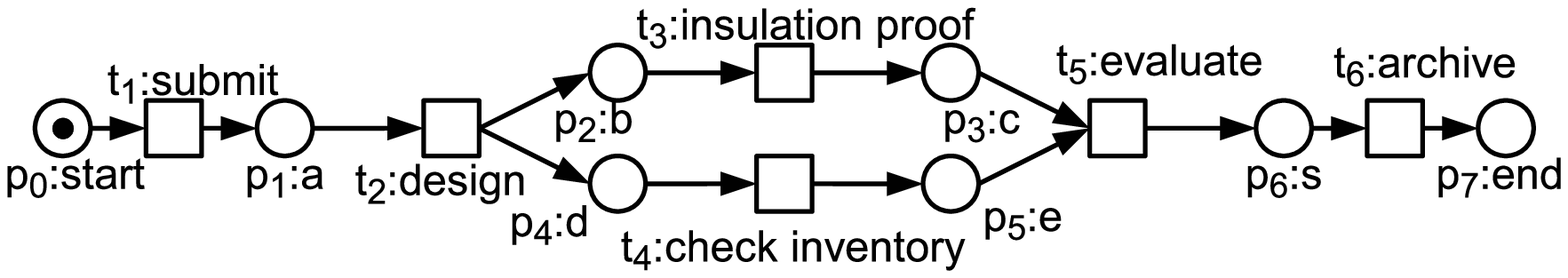}\\
(c) Representing execution as causal net\\
\caption{Example of conformance}
\label{fig-constraints-repair}
\end{figure}

Consider the execution trace in Figure \ref{fig-constraints-repair}(b) over the specification in Figure \ref{fig-constraints-repair}(a) from \cite{DBLP:conf/icde/0001SLZP15}.
We represent the corresponding causal net in Figure \ref{fig-constraints-repair}(c) as follows. 
For the first $t_1$ without any prerequisite, 
we put a place $p_0$ with $\pi(p_0)$ = start as the $pre$ set. 
The second $\sigma(2)$ of $t_2$ has prerequisite $pre_{F_\sigma} (pre_{F_\sigma} (t_2)) = {t_1}$.
We recover the labeling of the place $p_1$ between $t_2$ and its prerequisite $t_1$ to the place between $\pi(t_2)$ and $\pi(t_1)$ in the specification, i.e., $\pi(p_1) = a$. 
Similarly, considering the prerequisites of $t_5$, $pre_{F_\sigma} (pre_{F_\sigma} (t_5)) = {t_3, t_4}$, 
we obtain $\pi(p_3) = c, \pi(p_5) = e$. 
For the last $t_6$, which is not prerequisite of any others, a place $p_7$ : end is appended as $post_{F_\sigma} (t_6)$.

Referring to conformance definition, for any transition, say $t_1$ for instance, 
we have $\pi(pre_{F_\sigma} (t_1)) = \pi(p_0) = {start} = pre_{F_s} (submit) = pre_{F_s} (\pi(t_1))$ and $\pi(post_{F_\sigma} (t_2)) = \pi({p_2, p_4}) = {b,d} = post_{F_s} (design) = post_{F_s} (\pi(t_2))$.

\section{Approximate Pattern Matching}

Pattern queries are widely used in complex event processing (CEP) systems to discover situations of interest \cite{DBLP:conf/debs/MargaraCT14}.
However, current approaches cannot handle well the scenarios of approximation event pattern matching over extremely heterogeneous sources.
Zhang \cite{DBLP:journals/pvldb/ZhangDI10} studies pattern query in streams with imprecise occurrence times of events. 
However, it does not consider the heterogeneous problem of event data from difference sources.
Owing to the aforesaid heterogeneity and data quality issues, consistently answering the pattern queries \cite{DBLP:conf/sigmod/LianCS10} is unlikely.
A survey about complex event recognition (CER) techniques \cite{DBLP:journals/csur/AlevizosSAP17} summarize CER techniques that can handle incomplete and erroneous data streams or imperfect complex event patterns. 
Automata based techniques show limited or no support for pattern uncertainty.
First-order logic and probabilistic graphical models based techniques are harder to express data uncertainty.
Petri nets and grammars based techniques use implicit time representation without taking into account time itself as a variable.
Besides, all the techniques rely on experts to manually the weights/probabilities of the complex event patterns, which is hard in the scenarios of heterogeneous sources.
Considering the unpredictable order errors and even missing events in the sequence, as well as uncertainty in the pattern itself,
Li and Ge \cite{DBLP:conf/icde/LiG15} propose an approximate interleaving event matching method over sequence.

\section{Conclusion}

Event data are characterized by data elements being a function of time. In general, the datum takes the following form:
$D=\{(t_1, y_1), (t_2, y_2), \dots, (t_n, y_n)\}$
where $t_i, 1 \leq i \leq n$ are time stamps and $y_i = f(t_i)$ are data values.
As a special case of event data, time series data is important in IoT scenarios, 
where there are all kinds of sensor devices capturing data from physical world uninterruptedly.
The sensor devices are often unreliable and produce missed readings and unreliable readings \cite{DBLP:conf/pervasive/JefferyAFHW06}.
As a result, time series data are often large, incomplete \cite{DBLP:journals/pvldb/0001SZL13,DBLP:journals/tkde/0001SZLS16}
and dirty \cite{DBLP:conf/icde/0001SLZP15,DBLP:conf/sigmod/ZhangSW16}.
While 
sequential dependencies
\cite{DBLP:journals/pvldb/GolabKKSS09}
and 
speed constraints
\cite{DBLP:conf/sigmod/SongZWY15}
are some attempts of declaring constraints on consecutive data, 
more advanced studies are expected, 
e.g., to capture unique data quality challenges due to the presence of autocorrelations, trends, seasonality, and gaps in the time series data \cite{DBLP:journals/debu/DasuDS16}.
An important application of such techniques would be speech recognition,
since noises \cite{DBLP:journals/taslp/ZhangCF10,DBLP:journals/taslp/ZhangF12} and errors \cite{DBLP:journals/tslp/ZhangF12} affect significantly the speech summarization. 
It is promising to clean the data under various constraints before/during the speech recognition.

Consistency query answering over the inconsistent data has been studied in databases \cite{DBLP:conf/sigmod/LianCS10}, 
it is interesting to study the query answering over the event data. 
Our preliminary study in this topic \cite{DBLP:conf/cikm/Huang20} propose a method to conduct approximate event pattern matching over heterogeneous and dirty sources.

Finally, 
just like performing data mining and cleaning together shows better performance \cite{DBLP:conf/kdd/SongLZ15}, 
it is interesting to simultaneously cleaning the event data and mining process models,
since some intermediate errors can be noticed.

\bibliographystyle{abbrv}
\bibliography{event}

\end{document}